\documentclass{raa}        
\usepackage{graphicx,times}
\usepackage{natbib}
\usepackage{amssymb,amsmath}
\bibpunct{(}{)}{;}{a}{}{,}

\usepackage[driverfallback=dvipdfm]{hyperref}
\hypersetup{pdftitle = The title of my PDF, pdfauthor = My name, pdfsubject= The subject, pdfkeywords = keyword1 keyword2 keyword3} 
\hypersetup{colorlinks = true, linkcolor = green, anchorcolor = red, citecolor = blue, filecolor = red, pagecolor = red, urlcolor = red}

\begin{document}

   \title{Does ADS~9346 have a low-mass companion?
$^*$
\footnotetext{\small $*$ Supported by the Russian Foundation of Basic Research (projects 19-02-00843~A and 20-02-00563~A).}
}

 \volnopage{ {\bf 20XX} Vol.\ {\bf X} No. {\bf XX}, 000--000}
   \setcounter{page}{1}

   \author{O.V.~Kiyaeva\inst{1}, M.Yu.~Khovritchev\inst{1}, A.M.~Kulikova\inst{1}, N.V.~Narizhnaya
      \inst{1},  T.A.~Vasilyeva\inst{1}, A.A.~Apetyan\inst{1}
   }

   \institute{ Central Astronomical Observatory, Russian Academy of Sciences, 65/1 Pulkovskoye Chaussee, St.~Petersburg, 196140, Russia; {\it deimos@gaoran.ru}\\
%         \and
%              Yunnan Astronomical Observatory, National Astronomical Observatories, Chinese Academy of Sciences,
%              Kunmin 650011, China\\
% 	\and
% %	  Center for Astrophysics, University of Science and Technology of China, Hefei 230026, China\\
% Key Laboratory for Research in Galaxies and Cosmology, The University of Science
% and Technology of China, Chinese Academy of Sciences, Hefei, Anhui, 230026, China\\
% \and 
% Polar Research Institute of China,
% Jinqiao Rd. 451, Shanghai, 200136, China\\
\vs \no
   {\small Received 20XX Month Day; accepted 20XX Month Day}
}

\abstract{Based on the photographic and the CCD observations of the relative motion of A, B components of the binary system ADS~9346 obtained with the 26-inch refractor of the Pulkovo Observatory during 1979-2019, we discover an invisible companion associated with the A-star. Comparison of the ephemerides with the positional and the spectroscopic observations allowed us to calculate the preliminary orbit of the photocenter ($P=15$ years). The minimal mass of the companion is approximately  $0.13~M_\odot$. The existence of the invisible low-mass companion is implied by the IR-excess based on the IRAS data. To confirm this, additional observations of the radial velocity near the periastron need to be carried out.
}

   \authorrunning{Kiyaeva et al. }            %author_head in even pages
   \titlerunning{Does ADS~9346 have a low-mass companion?}  % title_head in odd pages
   \maketitle

%________________________________________________ sections below
% 
\section{Introduction}           %% first-level sections will be auto-capitalized
\label{sect:intro}
We have to take binary and multiple systems into account when we study the formation of stars and exoplanets and the further dynamical evolution of stellar groups and systems of exoplanets. This fact is reflected in many works on this topic (i.e.~\citealt{2014prpl.conf..267R}). That is why the search for multiple stars is a relevant observational problem, especially in the context of studies on exoplanet systems (i.e.~\citealt{2019MNRAS.490.5088M}). The search for binary and multiple stars is a traditional topic for Pulkovo Observatory (i.e.~\citealt{2006Ap.....49..397G}). We present the results of the study on the visual binary star ADS~9346 (WDS~14410+5757, HD~129580, HIP~71782), which has been a part of the 26-inch Pulkovo refractor observational program since 1979.

This paper continues the series of studies on ADS~9346 (\citealt{2008AstL...34..405K, 2010AstL...36..204K}). In the first work, based on all observations from the WDS catalog, the radius of the curvature of the observed arc and the dynamical mass of the binary star with the elliptical orbit were calculated. The value of the dynamical mass ($4.2\pm1.6\,M_\odot$) is significantly different from the expected according to the photometric data ($\approx 2.2\,M_\odot$). In the second work, using the Apparent Motion Parameters (AMP) method, we obtained three possible orbits with small inclination to the plane of projection and orbital periods $\approx 2000$ years. However, to achieve agreement with the expected mass we had to increase the parallax from Hipparcos ($18.9\pm0.9$ mas, ESA 1997) up to 24~mas. Also, based on the Pulkovo photographic (1979-2005) and CCD (2003-2007) observations, the perturbation in separation between the components was discovered and the assumption that the system might have another companion with the orbital period of 4 years (or as a multiple of 4). 

Nowadays new data became available: the series of Pulkovo CCD observations have been extended for 12 years; the parallaxes from Gaia~DR2~\citep{2018A&A...616A...1G} ($17.8573 \pm 0.0318$~mas for A-star and $17.8652 \pm 0.0384$~mas for B-star) and Gaia~EDR3~\citep{2021A&A...649A...1G} ($17.8383 \pm 0.0161$~mas for A-star and $17.8720 \pm 0.0429$~mas for B-star) are even smaller than parallaxes published in Hipparcos. Therefore, we have to take another look at our previous results.

In this work we study the discovered perturbations in the relative motion and define the preliminary inner orbit of the possible companion.

\section{Observational data analysis}
\label{sect:obs_data_analysis}
Our study is based on the CCD observations (77 series) obtained with the 26-inch refractor of the Pulkovo Observatory during 2003-2019~(\citealt{2010AstL...36..349I, 2020AN....341..762I}), which are available online in the Pulkovo database\footnote{\url{http://izmccd.puldb.ru/vds.htm}} and the Strasbourg astronomical Data Center (CDS)\footnote{\url{https://cdsarc.unistra.fr/viz-bin/cat/J/AN/341/762}}.

\begin{figure} 
   \centering
   \includegraphics[width=\textwidth]{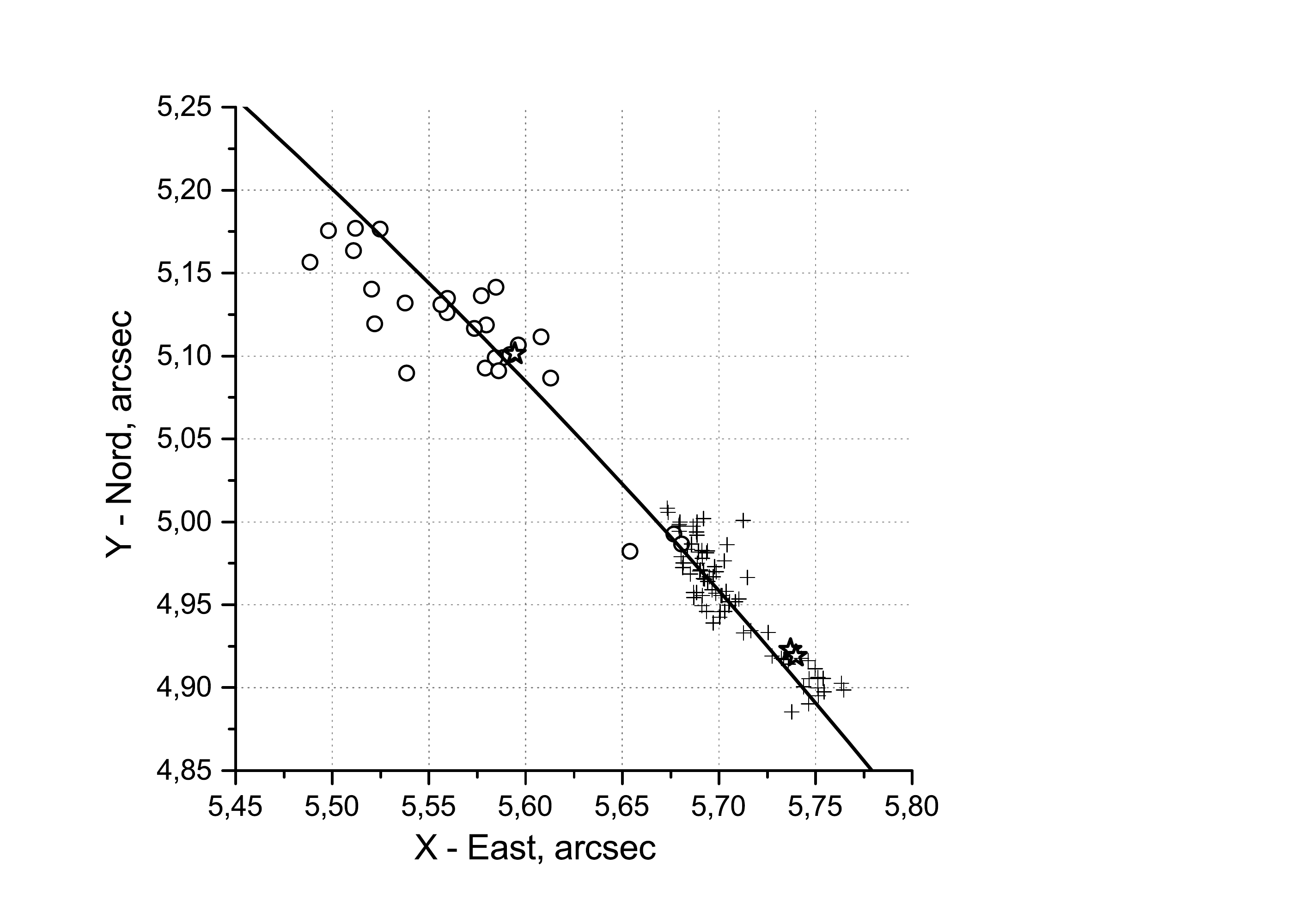}
   \caption{Observations obtained with the 26-inch refractor during 1979-2019: open circles --- photographic, crosses --- CCD,  open stars --- Hipparcos, Gaia~DR2 and EDR3, solid line --- the ephemeris of the orbit of the wide pair. } 
   \label{fig:pul}
   \end{figure}

The photographic observations from 1979 to 2005 (36 plates, 11 average annual positions) obtained with the same telescope add up to the CCD observations and at a first approximation reflect the orbital motion of the wide pair at this section of the orbit (\citealt{2014ARep...58...78K}). In the framework of this study we attempted to remeasure astronegatives which is described in section~\ref{sect:astronegative_remeasurements}. The series of Pulkovo observations from 1979 to 2019 is presented at Fig.~\ref{fig:pul}.

The results of 102 observations from 1830 to 2015 are presented in the WDS catalog~(\citealt{2016yCat....102026M}). The first observation by John Herschel is very different from the following measurements, so we will use the observation by Wilhelm Struve also obtained in 1830 as the first observation. Besides, we omit one observation from 1908 as it is an obvious mistake. We use observations from WDS to define more accurately the orbit of the wide pair obtained with the AMP method.

\begin{figure} 
   \centering
   \includegraphics[width=\textwidth]{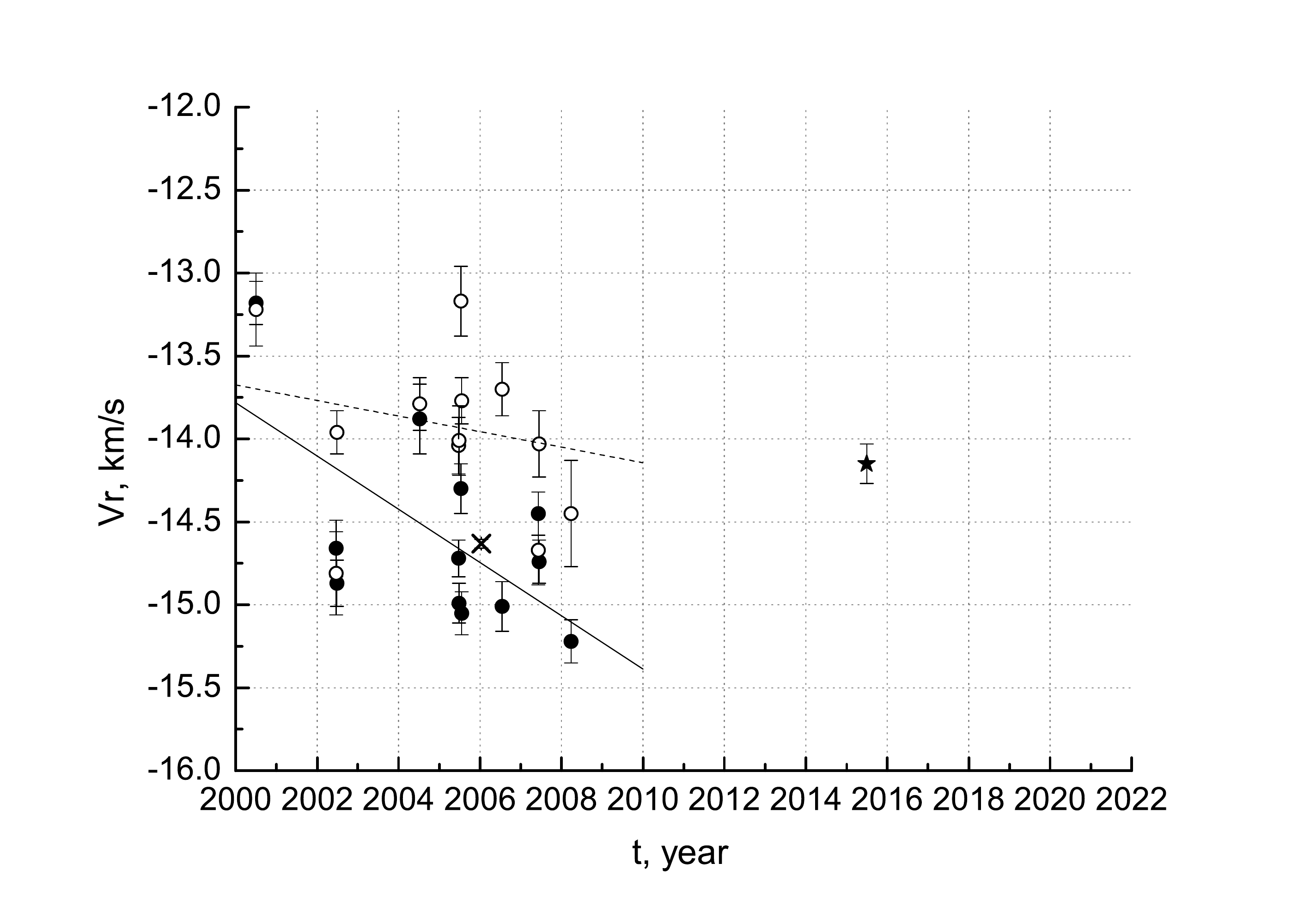}
   \caption{The radial velocity observations of the components: full circles -- the A-star velocity, open circles --- the B-star velocity, times sign -- Hobby-Eberly observation, a star -- Gaia observation. Lines show the linear trends for the A-star (solid line) and the B-star (dashes).} 
   \label{fig:VR}
   \end{figure}

Gaia~DR2 catalog~(\citealt{2018A&A...616A...1G}) contains observations of the radial velocity only for the A-star. According to CDS, the spectroscopic observations of the radial velocities for ADS~9346 are not presented in published works. However, there is one high-precision observation of the A-star obtained with the Hobby-Eberly telescope in 2006~(\citealt{2018A&A...615A..31D}). There are not any observations of the radial velocity of the component B in those works. The only long-term series of observations of both components is publushed in paper (\citealt{2010AstL...36..204K}). The velocities were measured in 2000-2008 with the CORAVEL -- a correlation radial velocity spectrometer constructed by A.~Tokovinin~(\citealt{1987SvA....31...98T}), mounted on the 1-m telescope of the Crimean Astrophysical Observatory. An assumption was made that according to the Tokovinin criteria~(\citealt{1988Ap.....28..173T}) the A-star might have a companion. Fig.~\ref{fig:VR} shows the results of those measurements and their dependency over time. Lines represent linear trends corresponding to the radial accelerations for each component on the given interval. The observations can not reflect the orbital motion of the external pair because the acceleration of the more massive component $\dot{V}_{rA}=-0.160\pm0.060$ km/s/year is greater than $\dot{V}_{rB}=-0.047\pm0.061$ km/s/year. This fact confirms the possibility of the A-star having an additional long-period companion. 

Gaia~DR2 contains the parallaxes of both components with the same accuracy ($0.03\,mas$). In the Gaia~EDR3 the error of the parallax is $0.016\,mas$ for the A-star and $0.043\,mas$ for the B-star. The parallax of the component B is measured with a lower accuracy and there are not any observations of the radial velocity. %This implies that the component B most likely has a short-period companion which was not discovered during rare radial velocity observations.

The fact that in the Gaia~EDR3 the parallax of the A-star is measured with a higher accuracy suggests that the orbital period of the possible companion should be significantly larger than the observational period of Gaia, the orbit should have a larger eccentricity, and during observations the companion should be located near the apoastron.

\section{Astronegative remeasurements}
\label{sect:astronegative_remeasurements}
As already mentioned in Introduction, the appearance of new high-precision data motivated us to conduct this research: more dense series of Pulkovo CCD observations, Gaia observations, radial velocity measurements. The methodology of astronegative measurements has also improved in the recent decade. That is why it seems reasonable to use new resources and repeat the digitization of astronegatives and the analysis of the scans to improve the accuracy of the photographic series for ADS~9346. The main goal of the procedure is to make sure that quasi-periodic perturbation that takes place for the CCD series is also present in old observations and is not a systematic effect associated with the CCD series only. 

Previously we used the flatbed scanner to digitize the photographic plates with the images of ADS~9346. The main imperfection of this scanner is the unstable field of systematic errors from one scanning to another (and even during the process of digitization). In the past few years, the MDD measuring complex has been actively used at the Pulkovo Observatory. (for detailed information see \citealt{2016AstL...42...41I, 2021A&A...645A..76K}).

\begin{figure} 
   \centering
   \includegraphics[width=\textwidth]{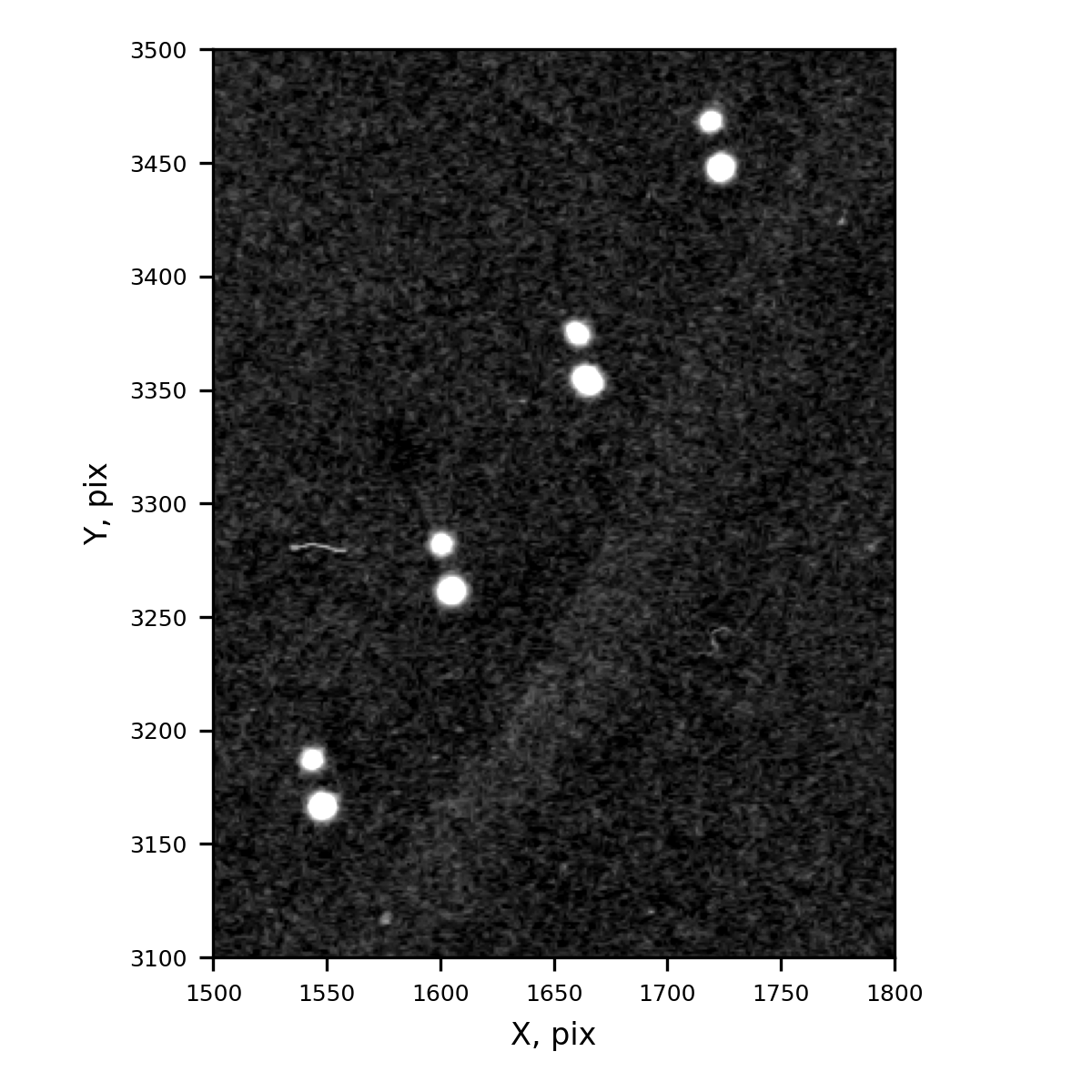}
   \caption{Images of the visual binary ADS~9346 at the fragment of the plate 10356 obtained with the 26-inch refractor of Pulkovo Observatory on May 13, 1979.} 
   \label{fig:plate10356}
   \end{figure}
   
\begin{figure} 
   \centering
   \includegraphics[width=\textwidth]{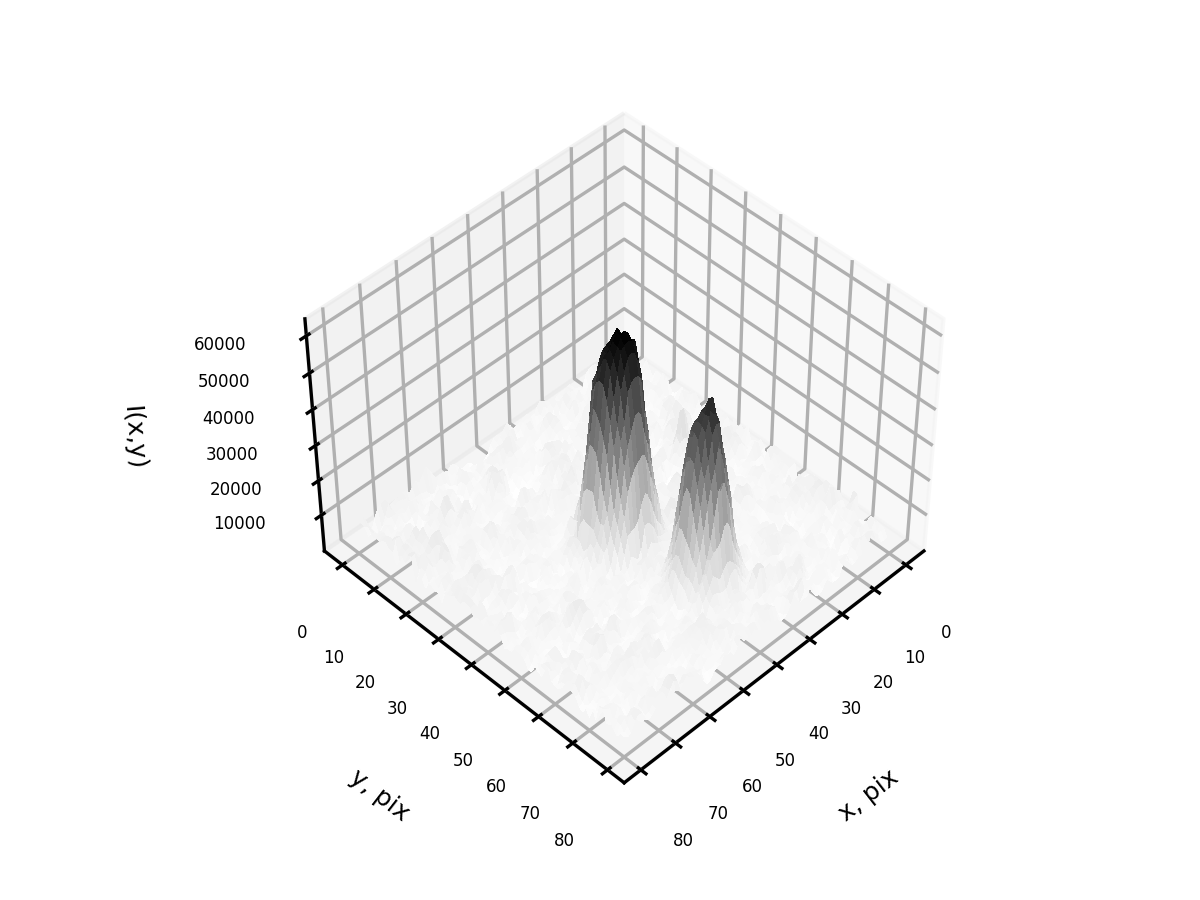}
  % \begin{minipage}[]{85mm}
   \caption{The image structure of ADS~9346 at the plate 10356.} 
%\end{minipage}
   \label{fig:plate10356_3D}
   \end{figure}

Figures \ref{fig:plate10356} and ~\ref{fig:plate10356_3D} show the fragment of the photoplate with the images of the ADS~9346 components and the structure of those images. This gives us a common understanding of the quality of astronegatives. All measurements were made using the technique described in the work~(\citealt{2021A&A...645A..76K}). To define the orientation and the scale we used reference stars which positions at the epoch of observations were taken from the Gaia~EDR3 catalog. In total we analyzed 37 astronegatives. The reference stars with the necessary signal to noise ratio were present at 27 from those astronegatives. Only these plates were analyzed later on. There were from 10 to 22 expositions taken on each plate, therefore we were able to estimate the intrinsic accuracy of the measurements: for $\rho$ -- $7\,mas$, for $\theta$ -- $0.05\,deg$. These estimates correspond to the level of accuracy of $1.5\,\mu m$ which is in agreement with the data from (\citealt{2021A&A...645A..76K}).

\begin{figure} 
   \centering
   \includegraphics[width=\textwidth]{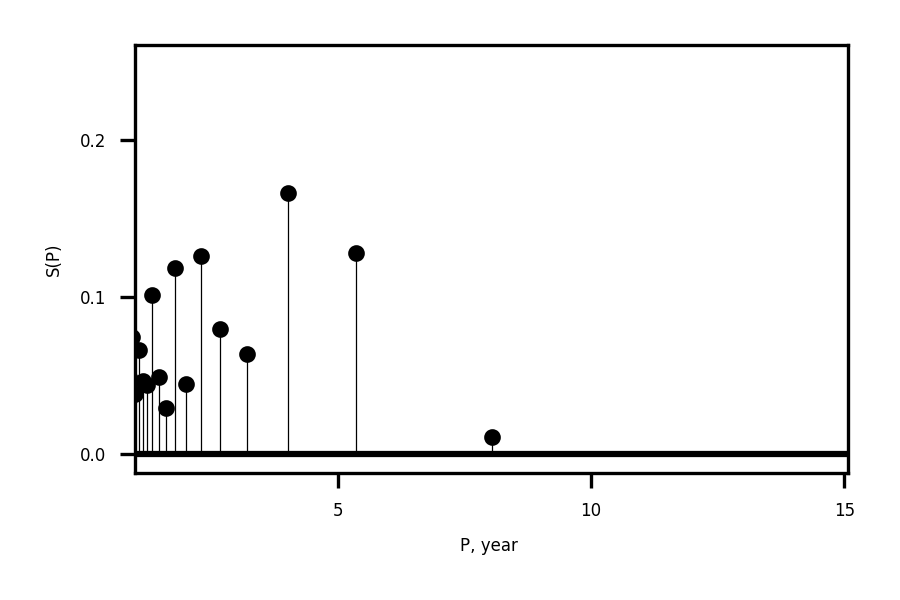}
   \includegraphics[width=\textwidth]{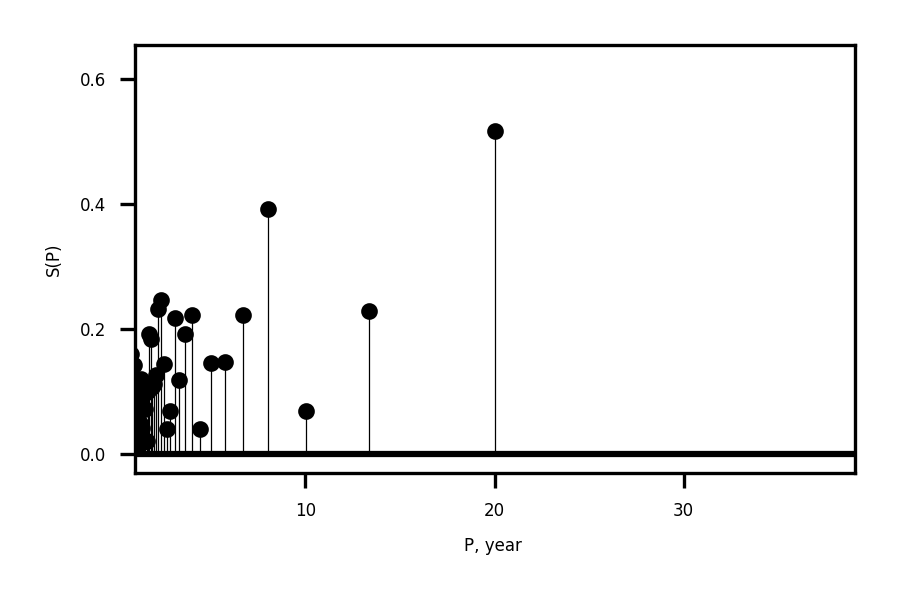}
   \caption{Power spectrums for the series of $\rho(t)$. Top plot takes into account only CCD observations. Bottom plot -- the whole series (photographic observations plus CCD observations).} 
   \label{fig:power_sp}
   \end{figure}

Fig.~\ref{fig:pul} represents the relative motion in the system ADS~9346. The initial series of photographic observations is systematically inconsistent with the series of CCD observations of $\rho$. It is difficult to come up with the reason of such divergence. The photographic series were obtained with the filter ZhS-18. The response curves of the photographic emulsion and the CCD sensor differ significantly. This could lead to a shift due to atmospheric dispersion, since the color indexes of the components are very different. Therefore, for the purposes of our research, the correction to the photographic series was determined empirically. The CCD series are shifted by $30\,mas$ for $\rho$, which corresponds to $1-\sigma$ for the photographic series. The series for $\rho(t)$ are characterized by noticeable quasi-periodic perturbations. They are more evident in CCD observations and can be noticed at a similar range in the photographic data. This allows us to assume that the attempt to repeat the digitization process was successful. The existence of variations greater than the standard error of $\rho$ suggests that this can be associated with the invisible low-mass companion in the system ADS~9346. More strictly it can be shown using the Fourier-analysis of the CCD series and the sum of the photographic and the CCD series. Fig. \ref{fig:power_sp} illustrates the power spectrums. As can be seen, using all observations we can obtain the period estimate close to 20 years. We should note that these series are quite irregular, therefore later this estimate will be improved in a more natural way for a given task.

\section{Photometric-based component masses}
\label{sect:photometric_masses}

\begin{figure} 
   \centering
   \includegraphics[width=\textwidth]{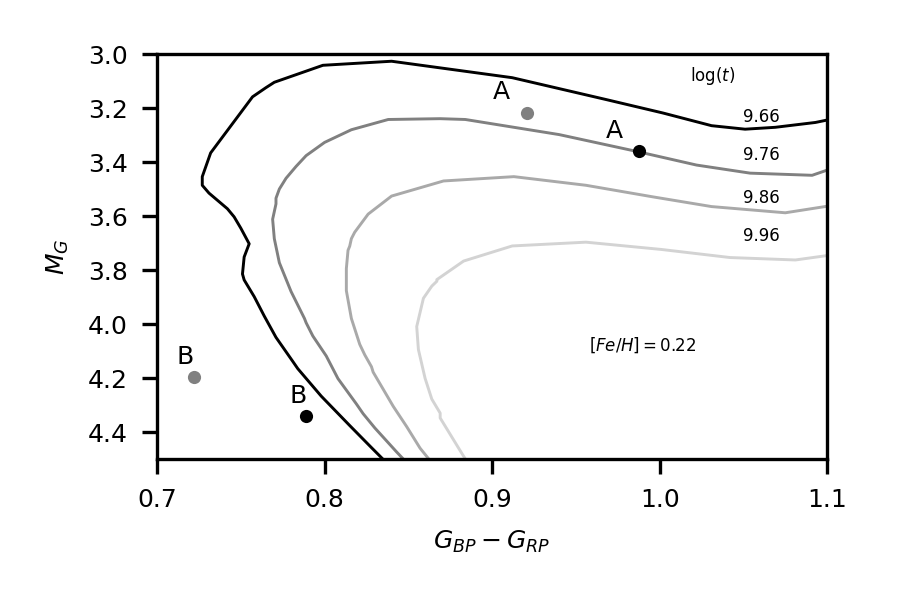}
   \caption{PARSEC-isochrones in the vicinity of ADS~9346 at the color-magnitude diagram corresponding to different ages and the metallicity $[Fe/H] = 0.22$ (taking the interstellar extinction and reddening into account). Black circles show the positions of the components according to the interstellar extinction from the works~(\citealt{2018A&A...615A..31D, 2019ApJ...877..116W}). Grey circles show the positions according to the Gaia~DR2 data.} 
   \label{fig:cmd}
   \end{figure}

The a priori mass estimate is one of the requirements of the AMP method. The high-precision photometric data allows us to hope for a sufficient accuracy of the mass determination. Taking into account the trigonometric parallax and the apparent magnitude estimates from Gaia~EDR3 alongside with the metallicity and the interstellar extinction data, we can easily put the components of ADS~9346 on the color-magnitude diagram. The PARSEC\footnote{\url{http://stev.oapd.inaf.it}}~(\citealt{2012MNRAS.427..127B}) tool makes it possible to plot the necessary isochrones (see Fig.~\ref{fig:cmd}). Then, we can interpolate and choose the isochrones and the points on those isochrones for the best fit in relation to the positions of the components A and B on the diagram.

As a result of the spectrum analysis during the radial velocity determination for the A-star (\citealt{2018A&A...615A..31D}), the values of the key parameters for this component were obtained. For instance, the metallicity $[Fe/H]=0.22$, the apparent color index $(B-V) = 0.9$~mag. The spectrum gives the intrinsic color index $(B-V)_0=0.86$~mag, meaning that the interstellar reddening $E(B-V)=0.04$~mag. Thus, we can easily estimate the interstellar extinction and reddening for the Gaia~EDR3 bands using the data from \citep{2019ApJ...877..116W}. The obtained values are the following: $A_G = 0.105$~mag, $E(G_{BP}-G_{RP}) = 0.055$~mag. On the other hand, there are direct measurements of the interstellar extinction and reddening in Gaia~DR2~(\citealt{2018A&A...616A...1G}): $A_G = 0.249$~mag and $E(G_{BP}-G_{RP}) = 0.123$~mag. The formal standard errors for these values do not exceed $0.01$~mag.

The results of the mass and age determinations for the A-star are presented in the work~(\citealt{2018A&A...615A..31D}): $m_A = 1.08~M_\odot$, $\log(t)=9.96$. These values were obtained using the technique similar to the one described in the first paragraph of this section. However, the authors based their research on the trigonometric parallaxes from Hipparcos2~(\citealt{2007A&A...474..653V}): $p_t=18.03\pm0.69$~mas. They suggest that the estimated value $p_t=22.2\pm4.23$~mas corresponds to the spectral data, which is not in agreement with the weighted average from Gaia~EDR3 ($p_t=17.86\pm0.04$~mas) within the limits of $1\sigma$. We suppose that the independent measurements from Hipparcos2 and Gaia are trustworthy, meaning that the mass determination needs to be reconsidered. The credibility of the results becomes debatable due to the uncertainties in the interstellar extinction and reddening estimates mentioned above (see Fig.~\ref{fig:cmd}). The middle ground can be found in the mass estimates about $1.21\pm0.1$~$M_\odot$ for the A-star and $1.05\pm0.15$~$M_\odot$ for the B-star. 

Thus, the analysis of the photometric and spectral data gives us 2.26~$M_\odot$ for the total mass and $\log(t)=9.76$ for age, which obviously corresponds better to the high metallicity. The age determination in the paper (\citealt{2018A&A...615A..31D}) seems too ambitious. A star that is older than 9~Gy with metallicity $[Fe/H] = 0.22$ is less likely to be located in the solar neighbourhood than a star with the same metallicity and the age that is a little over 5.7~Gy.

\section{Calculation of the orbital parameters of the photocenter of the A-star plus the invisible low-mass companion}
\label{sect:a-component-sat-orb}

The observational results obtained with the 26-inch refractor of the Pulkovo Observatory during 1979-2019 are presented at Fig.~\ref{fig:pul}. As mentioned above, there is a perturbation in the relative motion with the period around 20 years. To get more reliable orbital period estimates for the possible invisible companion, we added the Hipparcos and Gaia~DR2 data at the epoch 2015.5, Gaia~EDR3 data at the epoch 2016.0. The weights for observations were calculated according to the standard measurement errors. 

We can not determine which star has a companion using the relative positions, but the fact that the A-star being more massive also has a larger acceleration than the B-star gives us a reason to assume that only the A-star can have the long-period companion. Moreover, the list of stars with possible low-mass companions from the work~(\citealt{2019A&A...623A..72K}) contains the A-star, which is another confirmation of the existence of the companion.

The relative motion of the binary star with one companion can be expressed with the following equations:

\begin{equation}
\label{eq:one}
x_i=x_0+\dot{x}(t_i-t_0)+\ddot{x}(t_i-t_0)^2/2+BX_{i}+GY_{i}
\end{equation}
\begin{equation}
\label{eq:two}
y_i=y_0+\dot{y}(t_i-t_0)+\ddot{y}(t_i-t_0)^2/2+AX_{i}+FY_{i}
\end{equation}

Here $x_i=\rho\sin\theta$, \quad $y_i=\rho\cos\theta$ --- the apparent coordinates in the tangential plane, \quad $X_i=\cos{E_i}-e$, \quad $Y_i=\sqrt{1-e^2}\sin{E_i}$ --- 
the orbital coordinates calculated using the orbital period $P$, the periastron epoch $T$ and the eccentricity $e$ for each companion. $A$, $B$, $F$, $G$ --- the unknown Thiele-Innes elements which define the geometrical elements of the Kepler orbit of the photocenter $a_{ph}, i, \omega, \Omega$. The unknown parameters $x_0$, $\dot{x}$, $\ddot{x}$, $y_0$, $\dot{y}$, $\ddot{y}$ reflect the relative motion of the center of mass of the B-star in relation to the center of mass of the system A-star plus the invisible companion. Using these parameters, we can calculate the apparent motion parameters and the orbit of the external pair. 

First of all we considered periods in the range from 10 to 15 years with a step size of 1 year, because the series of CCD observations are 16 years long. $T$ and $e$ were obtained using the brute-force search method with the step size of 0.1 year for $T$ and 0.1 for $e$. The criteria of the method is the minimal standard deviation $\sigma_{xy}=\sqrt{\sigma_x^2+\sigma_y^2}$. Here $\sigma_x=\sqrt{\sum{p_i(dx_i)^2}/n}$, $p_i$ -- the weights corresponding to the observational error, $\sigma_y$ is calculated in the same way.

\begin{figure} 
   \centering
   \includegraphics[width=0.7\textwidth]{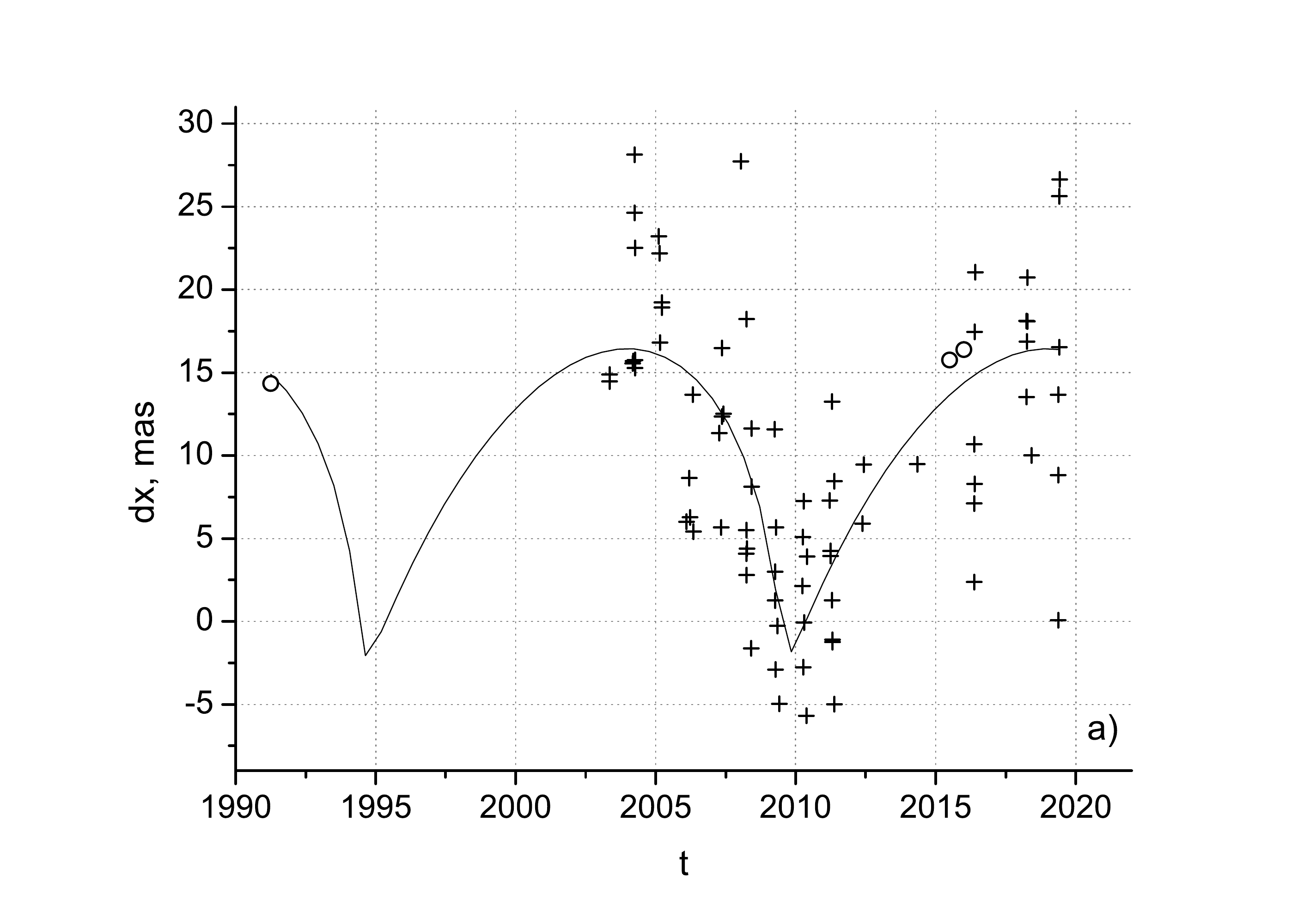}
   \includegraphics[width=0.7\textwidth]{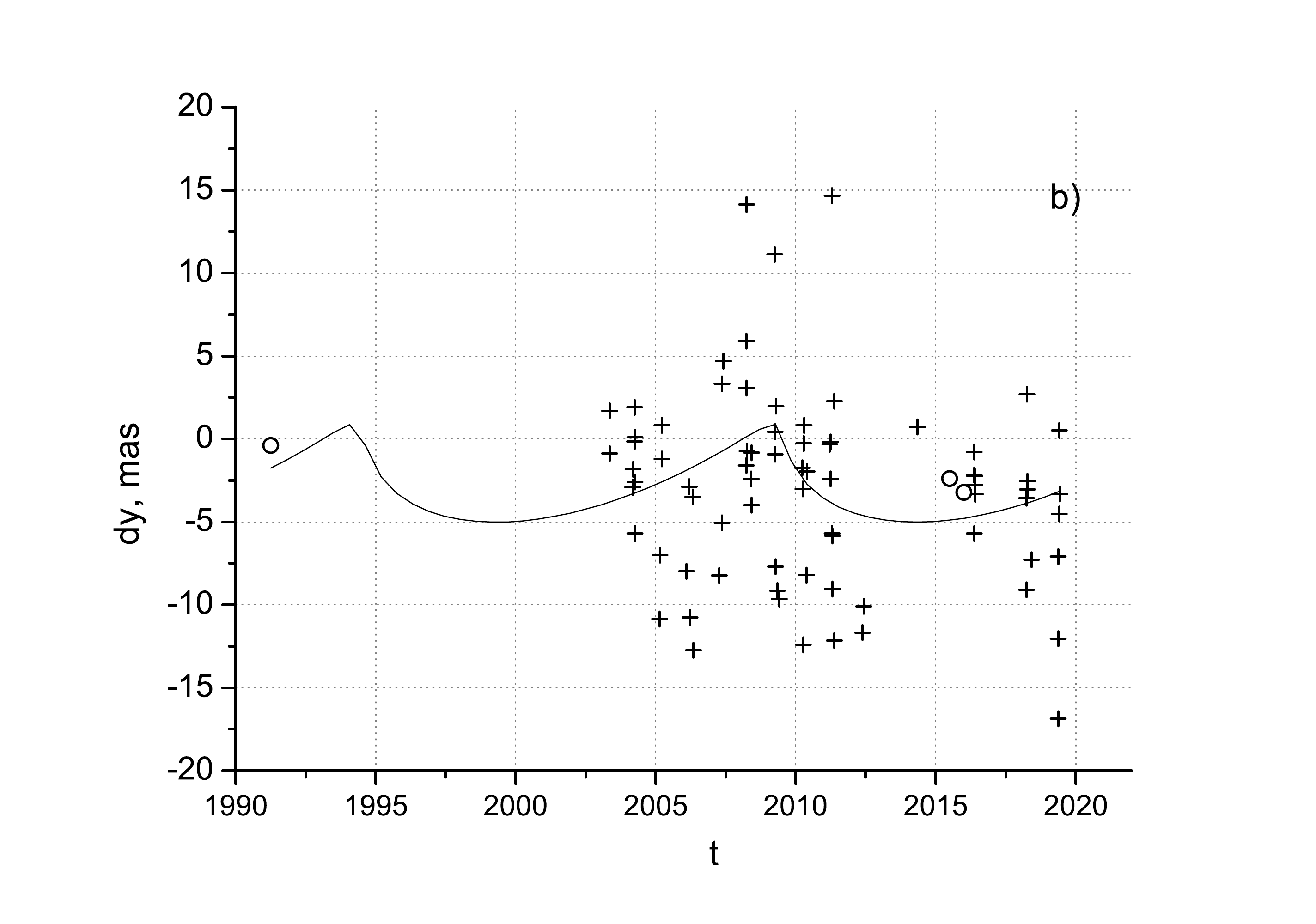}
   \includegraphics[width=0.7\textwidth]{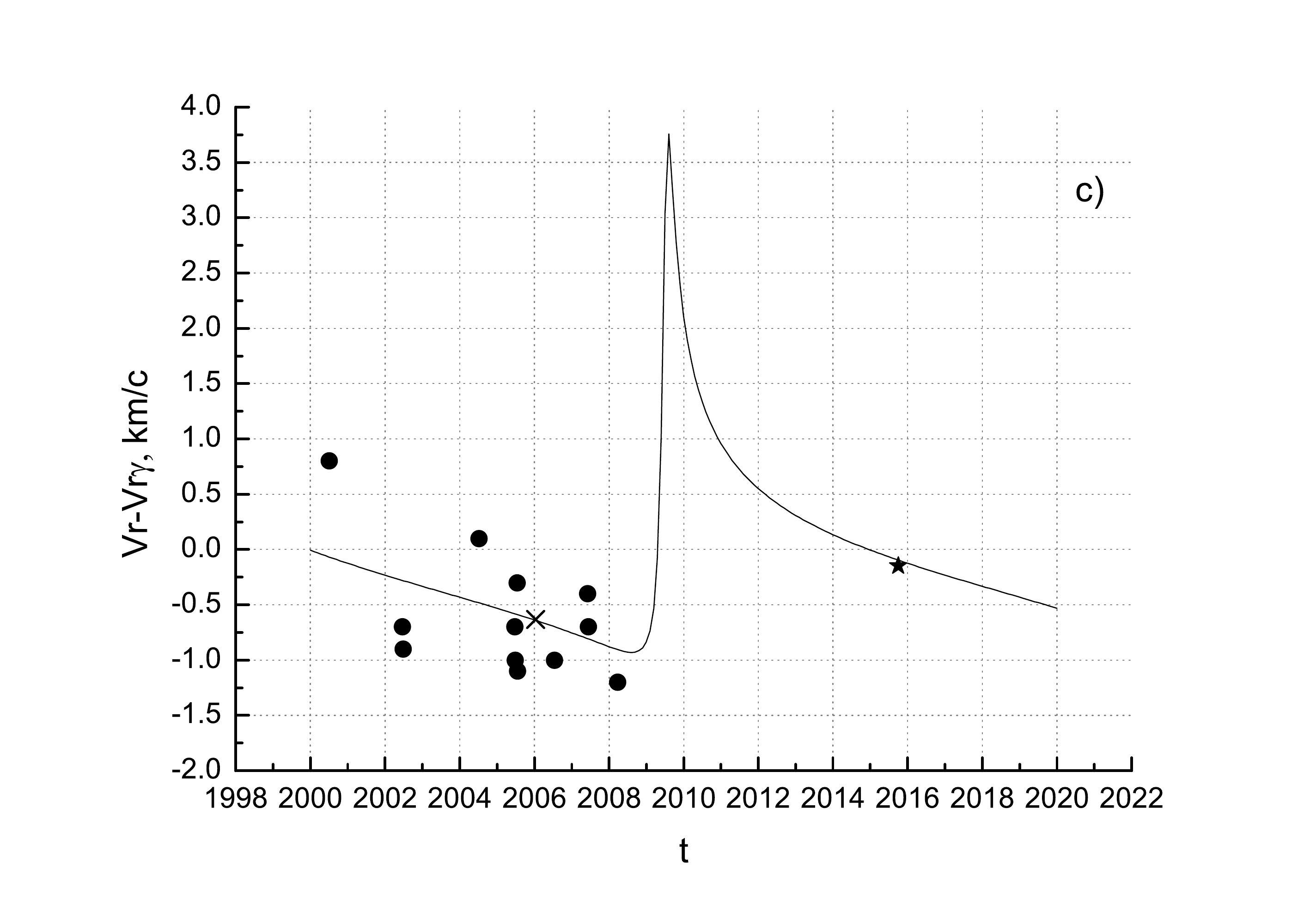}
   \caption{The comparison between the ephemerides and the positional observations of the companion in right ascension (a), declination (b) and the observed radial velocity (c). For the positional observations: crosses --- CCD observations, open circles --- Gaia and Hipparcos. For the radial velocity observations: full circles --- CORAVEL, times signs --- Hobby-Eberly observations, a star --- Gaia~DR2.}
   \label{fig:os_AB}
   \end{figure}

According to the given criteria, all orbits with the periods from 11 to 15 years fit the observational data equally good. Solving the equations \ref{eq:one} and \ref{eq:two} we can obtain the Thiele-Innes elements (that define the orbit of the companion) with great confidence. Hence, we use the independent observations of the radial velocities as a criteria: $t=2006.03, \Delta{V_{rA}}=V_{rA}-V_{rA\gamma}=-0.632\pm0.027$~km/s~\citep{2018A&A...615A..31D} and $t=2015.5, \Delta{V_{rA}}=-0.15\pm0.12$~km/s (Gaia~DR2), if we set $V_{rA\gamma}\approx-14.0$~km/s. (see Fig.~\ref{fig:os_AB}).

Additionally, we compare the trend obtained using the ephemerides from 2000-2008 with the observed radial velocity acceleration of the A-star ($-0.16\pm0.06$ km/s/year).

We get the best solution (an agreement within the confidence intervals) for the orbit with $P=15\pm0.5$~years, $e=0.86\pm0.01$, $T=2009.5\pm0.1$. The corresponding ephemerides: $t=2006.0, \Delta{V_{rA}}=-0.637$~km/s; $t=2015.5, \Delta{V_{rA}}=-0.07$~km/s; $\dot{V}_{rA}=-0.11$~km/s/year.

The solution of the equations \ref{eq:one} and \ref{eq:two}, as well as the radial velocity semi-amplitude of the visible component, the semi-major axis of the relative orbit and the elemnents of the Kepler orbit corresponding to the Thiele-Innes elements are presented in Table~\ref{tab:orb-ti}. 

\begin{table}
\bc
\begin{minipage}[]{100mm}
\caption[]{The parameters of the relative motion of the photocenter of A-star plus low-mass companion at  2010.0.\label{tab:orb-ti}}\end{minipage}
\setlength{\tabcolsep}{2.5pt}
\small
 \begin{tabular}{rr}
  \hline\noalign{\smallskip}
$P_{in}$, yr & $15\pm0.5$ \\

$x_0$, mas  &  $5695.1\pm 1.3 $ \\
$y_0$, mas  &  $4960.3\pm 1.5 $  \\
 $\dot{x}$, mas/yr & $5.07\pm0.14$  \\
 $\dot{y}$, mas/yr & $-6.63\pm0.17$  \\
 $\ddot{x}$, mas/yr$^2$ & $-0.1130\pm0.0302$  \\
 $\ddot{y}$, mas/yr$^2$ & $0.0968\pm0.0345$  \\
\hline\noalign{\smallskip}
$T$, yr &  $2009.5\pm0.1$ \\
$e$ & $0.86\pm0.01$ \\
\hline $B$, mas & $-8.33\pm1.30$ \\
$A$, mas & $2.39\pm1.51$ \\
$G$, mas & $-7.97\pm1.72$ \\
$F$, mas & $-3.41\pm2.01$ \\
\hline $a_{ph}$, mas & $11.55(-0.23)\pm2.46(2.45)$  \\
$i, ^\circ$ & $110.8(-15.8)\pm22.0(14.3)$ \\
$\omega, ^\circ$  & $314.7(+17.5)\pm32.1(28.6)$ \\
$\Omega, ^\circ$  & $266.1(+7.5)\pm21.1(19.9)$ \\
\hline\noalign{\smallskip}
$a$, a.e. & 7.0  \\
$K_{ph}$, km/s  & 2.4  \\
\hline\noalign{\smallskip}
$M_1, M_\odot$  & 1.21 \\
$m_s, M_\odot$  &  0.129 \\
\hline $\sigma_x$, mas & 9.20  \\
$\sigma_y$, mas & 9.39  \\
$\sigma_{xy}$, mas & 13.14 \\
  \noalign{\smallskip}\hline
\end{tabular}
\ec
%% place \tablecomments and \tablerefs below \end{center| and \end{center}:
%% you may leave the table-width parameter to editors or set to your actual size
\tablecomments{0.86\textwidth}{The Kepler elements correspond to the Thiele-Innes elements. The shift of the mean solution and its error are in brackets. $a_{ph}$ --- the semi-major axis of the photocentric orbit, $a$ --- the semi-major axis of the relative orbit, $K_{ph}$ --- the half range of the photocentric radial velocity. The minimal mass of the companion $m_s$ is calculated using the parallax from Gaia~EDR3 and the mass of the visible component($M_1$.)}

\end{table}

If we believe that the photocenter coincides with the visible component, we can calculate the minimal mass of the companion $m_s$ which depends on the parallax and the mass of the visible component $M_1$. Here we use the weighted average parallax of the A-star from Gaia~EDR3: $p_t = 17.838\pm0.016$~mas.

The residual errors of the Kepler orbital elements were obtained with the Monte-Karlo method in the following way. All residuals were randomly altered 50 times with the dispersion 9~mas which corresponds to the variations (see Tab.~\ref{tab:orb-ti}). We calculated the mean solution shifted in relation to the initial (model) solution. The errors were calculated in relation to the model solution. The shift and the errors of the mean solution are presented in brackets.

The graphical representation of the observations and the ephemerides of the right ascention $(dx)$, the declination $(dy)$ and the radial velocity ($\Delta{V_r}$) are shown at Fig.~\ref{fig:os_AB}.

There is a systematic difference between space observations and the observations with the 26-inch refractor which is significant for some stars. For this star the difference between the Gaia~DR2 system and the CCD observations is 6~mas for $\rho$. Therefore, it is worth mentioning that at the $dx(t)$ plot the changes in the Gaia~DR2 (2015.5) and the Gaia~EDR3 positions are in full agreement with the ephemerides for this area.

We can conclude that the effect is significant ($a\approx11$~mas), the inclination of the orbit is close to $90^\circ$, in the tangential plane the companion moves in the line of the orbital node  ($\Omega\approx270^\circ$), which is in agreement with the $dx(t)$ plot. There is still an unaccounted effect with the possible period of 4 years at the $dy(t)$ plot.

\begin{figure} 
   \centering
   \includegraphics[width=\textwidth]{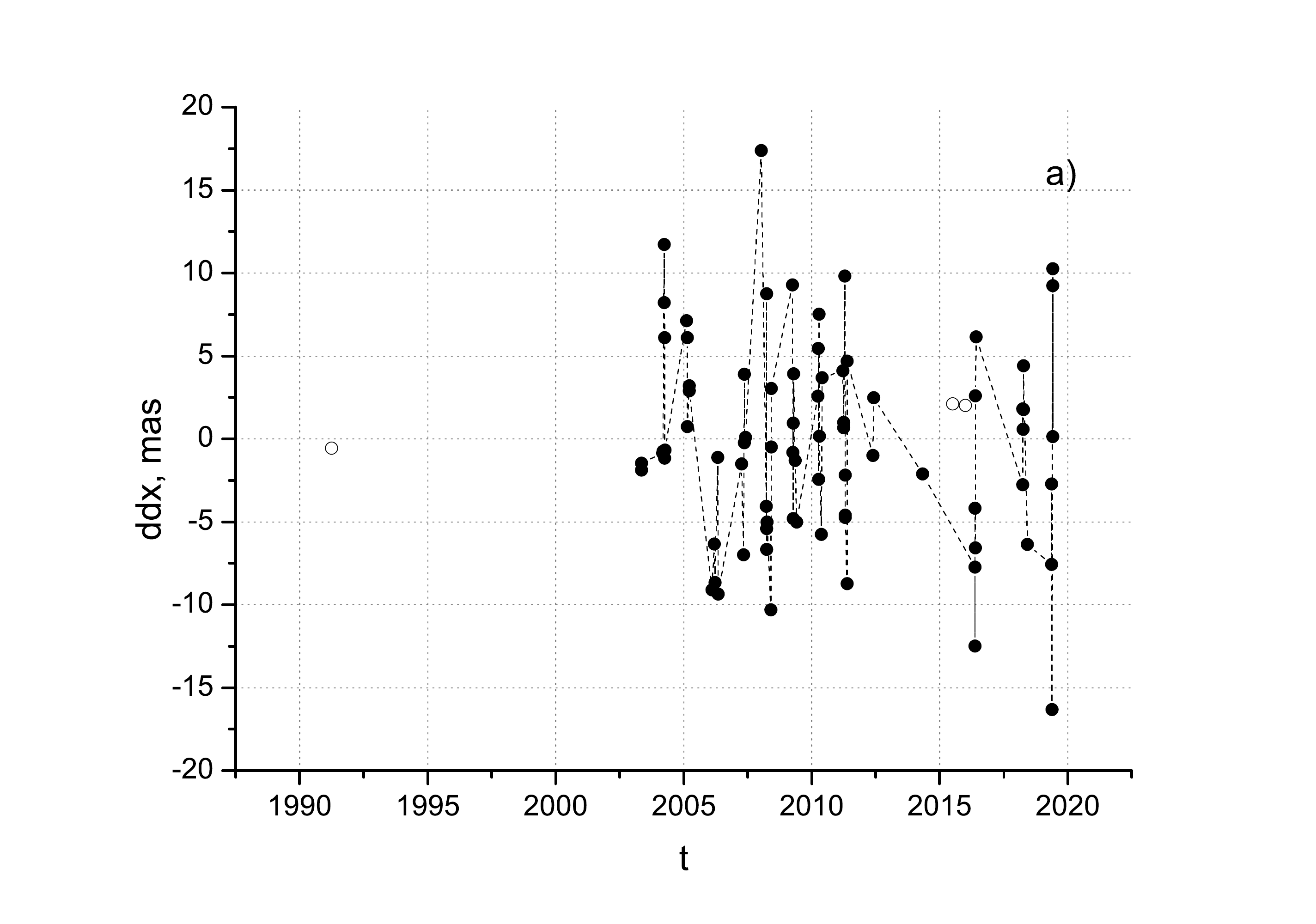}
   \includegraphics[width=\textwidth]{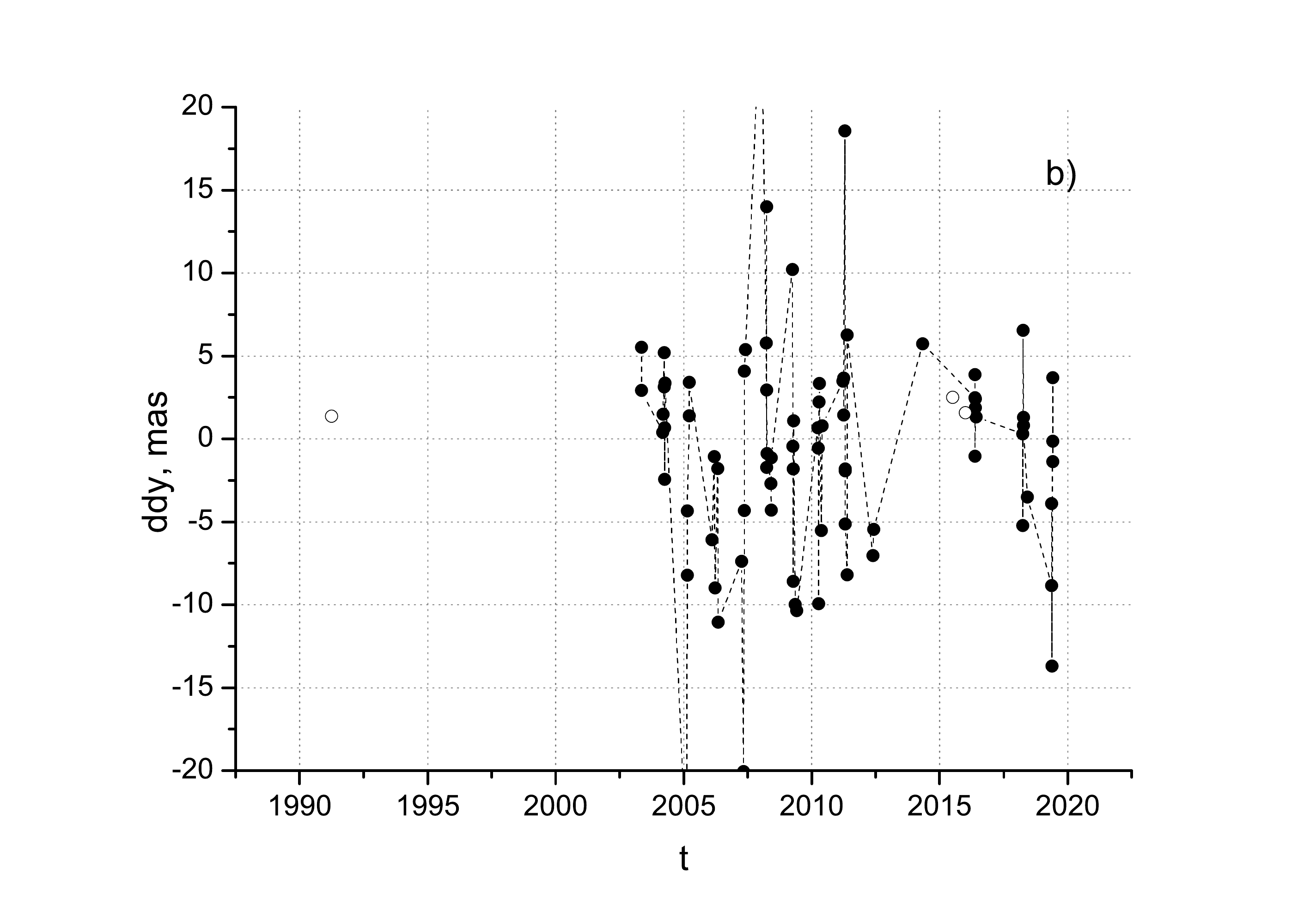}
  % \begin{minipage}[]{85mm}
   \caption{The residuals (after taking into account the companion) of the right ascension (a) and the declination (b). Full circles --- CCD observations, open circles --- the average annual positions of the photographic observations, stars --- the Hipparcos and Gaia data.} 
%\end{minipage}
   \label{fig:dev}
   \end{figure}

Fig.~\ref{fig:dev} shows the residuals after taking into account the companion with the orbital period of 15 years.

The calculated orbit with the period of 15 years can be improved or changed when the radial velocity observations in the area of periastron become available.

\section{Estimation of the values of the AB pair orbital elements}
\label{sect:a-b-pair-orb}

One of the the best ways to determine the initial orbit of a visual binary using the short arc in the framework of the two-body problem is the AMP method. It was described numerous times starting from 1980 (i.e. \citealt{1980AZh....57.1227K, 2020AstL...46..555K}). After taking the motion of the companion into account (see Tab.~\ref{tab:orb-ti}) we calculated the apparent motion parameters at $t_0=2010.0$ (the separation between the components ($\rho$) in arcseconds, the positional angle ($\theta$) in degrees, the apparent relative motion ($\mu$) in mas per year and the positional angle of the apparent motion ($\psi$) in degrees) and their errors using the following simple expressions:

$$\rho=\sqrt{x^2+y^2}~, \qquad \theta=\arctan\frac{x}{y},$$
$$\mu=\sqrt{\dot{x}^2+\dot{y}^2}~, \qquad \psi=\arctan\frac{\dot{x}}{\dot{y}}.$$
$$\varepsilon_{\rho}=\sqrt{(x\varepsilon_{x})^2+(y\varepsilon_{y})^2}/\rho~, \qquad
\varepsilon_{\theta}=\sqrt{(y\varepsilon_{x})^2+(x\varepsilon_{y})^2}/\rho^2
$$
$$\varepsilon_{\mu}=\sqrt{(\dot{x}\varepsilon_{\dot{x}})^2+(\dot{y}\varepsilon_{\dot{y}})^2}/\mu~, \qquad
\varepsilon_{\psi}=\sqrt{(\dot{y}\varepsilon_{\dot{x}})^2+(\dot{x}\varepsilon_{\dot{y}})^2}/\mu^2
$$

In addition, we also obtain from the observation the parallax, the relative radial velocity and the total mass of the system.

We use the weighted average of the parallax $p_t$ from Gaia~EDR3, which is $17.842\pm0.046$~mas.

We set $V_{rA\gamma}=-14$km/s. Then, according to the radial velocity observations during 2000--2008, the relative radial velocity $V_{rB}-V_{rA}\approx0\pm0.5$~km/s. For a star with the orbital period about several thousands of years this value will not change significantly in 10 years. 

The radius of the curvature $\rho_c$ can be calculated using the following equation:
\begin{equation}
\label{eq:three}
\rho_c=|\frac{(\dot{x}^{2}+\dot{y}^2)^{1.5}}{\dot{x}\ddot{y}-\dot{y}\ddot{x}}|
\end{equation}
However, the determination of $\rho_c$ with this equation (\ref{eq:three}) is associated with uncertainties due to large errors $\ddot{x}$ and $\ddot{y}$ ($\rho_c=2.2''\pm4.0''$), as a result we can not determine the distance between the components $r$ at the moment $t_0$ using the equation~\ref{eq:four}:
\begin{equation}
\label{eq:four}
r^3=4\pi^2(M_A+M_B)\frac{\rho\rho_c}{\mu^2}\mid{\sin(\psi-\theta)}\mid
\end{equation}

That is why for all variations we get the set of elliptical orbits which satisfy the following condition:

\begin{equation}
\label{eq:five} \frac{\rho}{p_t}\leq{r}\leq\frac{8\pi^2}{v^2}(M_A+M_B)
\end{equation}

Here $v$ --- the space velocity of B-star in relation to A-star.

From this set of orbit we choose the best fit for all observations of the binary star from its discovery. This solution is measured by the angle $$\beta=\pm\arccos\frac{\rho}{p_tr}$$ between the direction to the companion and its projection on the tangential plane, as was suggest in the paper~(\citealt{1996ARep...40..795K}).

The total mass of all visible components and the companion equals $2.26~M_\odot$. Taking into account the influence of the possible companion, we suppose that the total mass of the system is unknown. Then we correct the value using all available observations. When choosing the best fit, we do not compare the observations directly, but use the agreement between the Thiele-Innes elements (A, B, F, G) instead. They can be calculated using the geometrical orbital elements ($a, i, \omega, \Omega$) without observations, and also using the dynamical elements ($P, T, e$) combined with observations separated in time (\citealt{1983AZh....60.1208K}). The criteria for this method is the minimum of the function: 
\begin{equation}
\label{eq:six} S=\sqrt{\Delta{A}^2+\Delta{B}^2+\Delta{F}^2+\Delta{G}^2}
\end{equation}

Here $\Delta{A}, \Delta{B}, \Delta{F}, \Delta{G}$ --- the differences between the Thiele-Innes elements obtained in two ways. As opposed to the direct comparison between the observations and the ephemerides, in this case we do not need to assign weights to particular heterogeneous observations which inevitably bring in some subjectivity. However, it is important to have several reliable points spaced along the entire arc near the middle of the observational sector. 

As the range of observations from the WDS catalog is too wide, we calculated the additional reference series through the middle of the observational sector. All observations before 1991 (Hipparcos) were split in 30-40~years intervals and the apparent motion parameters ($\rho, \theta$) were calculated for the mean moment of each interval. Besides, the reference series include the first observation conducted by Struve, Hipparcos and Gaia data, and the AMP obtained using the photographic Pulkovo observations. Comparing the AMP-orbits with the reference series, we get the best formal solution.

The results --- AMP, $\beta_1$, and the total mass  $M_{A+B}$, which correspond to the minimum of $S_1$, $\beta_2$ and $S_2$ with the fixed mass $M_{A+B}=2.4M_\odot$ --- are presented in Tab.~\ref{tab:pvd}. Two solutions are shown: the AMP obtained using the equations \ref{eq:one} and \ref{eq:two} (see Tab.~\ref{tab:orb-ti}) taking into account the companion with the period 15~years (solution 1) and homogeneous CCD observations without the companion (solution 2).

\begin{table}
\bc
\begin{minipage}[]{100mm}
\caption[]{The parameters of the relative motion of the A-B pair, calculated for different conditions\label{tab:pvd}}\end{minipage}
\setlength{\tabcolsep}{2.5pt}
\small
 \begin{tabular}{ccrrrrrccrc}
  \hline\noalign{\smallskip}
Var. & $t_0$ & $\rho, ''$ & $\theta, ^\circ$ & $\mu, ''$/yr
& $\psi, ^\circ$ & $\beta_1, ^\circ$ &  $M_{A+B}, M_\odot$ &
$S_1,  ''$  & $\beta_2, ^\circ$ &  $S_2,  ''$ \\
\hline\noalign{\smallskip}
 1 & 2010.0     & 7.5524 & 48.945 & 0.0083 & 142.6 & 0 & 4.4 & 0.175 & 0 & 0.524 \\
   &         & $\pm0.0014$ & 0.010 & 0.0002 & 1.0 & 10 &  &  & 6 & \\
\hline\noalign{\smallskip}
  2 & 2010.0 & 7.5560 & 48.989 & 0.0085 & 141.5 & $\pm14$ & 4.1 &  0.098 & 0 & 0.381 \\
 & & $\pm0.0008$ & 0.004 & 0.0001 & 1.3 & 7 &  &  & 10 & \\
  \noalign{\smallskip}\hline
\end{tabular}
\ec
\tablecomments{0.86\textwidth}{Two solutions are shown: 1 -- using the CCD observations with the companion ($P_{in}=15$~years); 2 --  using the CCD observations without the companion. The best fit corresponds to the minimum of $S_1$, if the mass
$M_{A+B}$ is undefined, or $S_2$, if $M_{A+B}=2.4~M_\odot$.}
\end{table}

\begin{table}
\bc
\begin{minipage}[]{100mm}
\caption[]{The orbital parameters of the A-B pair\label{tab:orbout}}\end{minipage}
\setlength{\tabcolsep}{2.5pt}
\small
 \begin{tabular}{crrrrrrr}
  \hline\noalign{\smallskip}
ref
 &\multicolumn{2}{c}{This work,}  & \multicolumn{2}{c}{This work,}
 & \multicolumn{2}{c}{\cite{2019AstL...45...30I}} & \cite{2010AstL...36..204K}\\
 &\multicolumn{2}{c}{ solution 1} & \multicolumn{2}{c}{solution 2}
 & \multicolumn{2}{c}{ } &  \\
\hline\noalign{\smallskip}
P, yr & 5435    & 2455 & 5725   & 2749 & 2547  & 4130 & 2644 \\
       &$\pm519$ &  213 &  556   & 220  & 1031  & 2230 & 2239\\
\hline\noalign{\smallskip}
T, yr & 3612    & 3165 & 3170   & 3323 &  1528 & 1526 & 3100\\
       &$\pm128$ &  103 &  599   &  814 &   398 & 954  & 2241\\
\hline\noalign{\smallskip}
e      &   0.07  & 0.45 &  0.05  & 0.39 & 0.75  & 0.80 & 0.34\\
       &$\pm.02$ &  .04 &   .04  &  .04 &  .20  &  .18 &  .22\\
\hline\noalign{\smallskip}
$a, ''$ &   7.38  & 5.24 & 7.64  & 5.56 &  11.48 & 15.57 & 5.78\\
        &$\pm.47$ &  .30 &  .49  &  .30 &   5.20 &  7.72 & 2.39\\
\hline\noalign{\smallskip}
$i, ^\circ$ & 0 &  0 &  0 & 14 & 71  &  74 & 19\\
      &$\pm 17$ & 20 & 16 &      8  & 11  &  11 &  8\\
\hline\noalign{\smallskip}
$\Omega+\omega, ^\circ$ &  162 & 224 & 127 & 225 &  - & - & 212\\
       & $\pm128$ & 86 & 126 & -  & - & - & 22 \\
\hline\noalign{\smallskip}
$\Omega, ^\circ$ & - & - & - & 142 &  16 & 18 & 159\\
       & $\pm$- & - & - & 43 &  38 & 61 & - \\
\hline\noalign{\smallskip}
$\omega, ^\circ$ & - & -  & - & 83  & 273 & 283 & 53\\
       &$\pm$  - & - & -  & 36 & 18 & 20 & - \\
\hline\noalign{\smallskip}
$M_{A+B}$, $M_\odot$ & 2.4 & 4.2  & 2.4 & 4.1  & 41 & 38 & 4.8 \\
\hline $\sigma_\rho$, mas & 113.5 & 112.4 & 112.8 & 112.3 &  114.0 & 112.6 & 122.2   \\
$\sigma_\tau$, mas & 131.2 & 131.1& 130.5 & 130.4 &  131.8 & 132.4 & 132.4   \\
  \noalign{\smallskip}\hline
\end{tabular}
\ec

\tablecomments{0.86\textwidth}{For orbits determined in this paper $\Delta{V_r}=0$~km/s is fixed. Total masses for all orbits correspond to the parallaxes from Gaia~EDR3 (17.8425~mas). $\omega$ and $\Omega$ can differ by $180^\circ$. Due to little inclinations of orbits in relation to the tangential plane, in 2010 we could obtain only $\omega+\Omega$. Now, using the Gaia data, we can separate these parameters.}
\end{table}

We obtained the orbital elements for both variants with the expected total mass $2.4\,M_\odot$ and the mass corresponding to the minimum of function S. The results are presented in Tab.~\ref{tab:orbout}.

For comparison we also show the orbits from \cite{2019AstL...45...30I} with the period 2547~years (without weights) and 4130~years (with weights), as well as the orbit obtained in the work~(\citealt{2010AstL...36..204K}) using the homogeneous photographic observations, AMP at the epoch 1990.0, for which the obtained total mass was also too high. Total masses for all orbits correspond to the parallaxes from Gaia~EDR3. Within the limits of errors we achieve a good agreement of all three AMP-orbits with the mass that exceeds $4 M_\odot$. 

The errors of the AMP-orbits are defined by the errors of the initial data. The errors of the elements obtained in 2010 are too large due to the large parallax error used back then.

Since $\Delta{V_r}\approx0.0$~km/s, the location of the node can be determined to $180^\odot$. Since $\beta$ is small, we can conclude that the plane of the inner orbit is oriented near the tangential plane. A small inclination is characteristic for all AMP-obits. With the expected mass  $2.4 M_\odot$ we get almost circular orbits in the tangential plane.

\begin{figure} 
   \centering
   \includegraphics[width=0.7\textwidth]{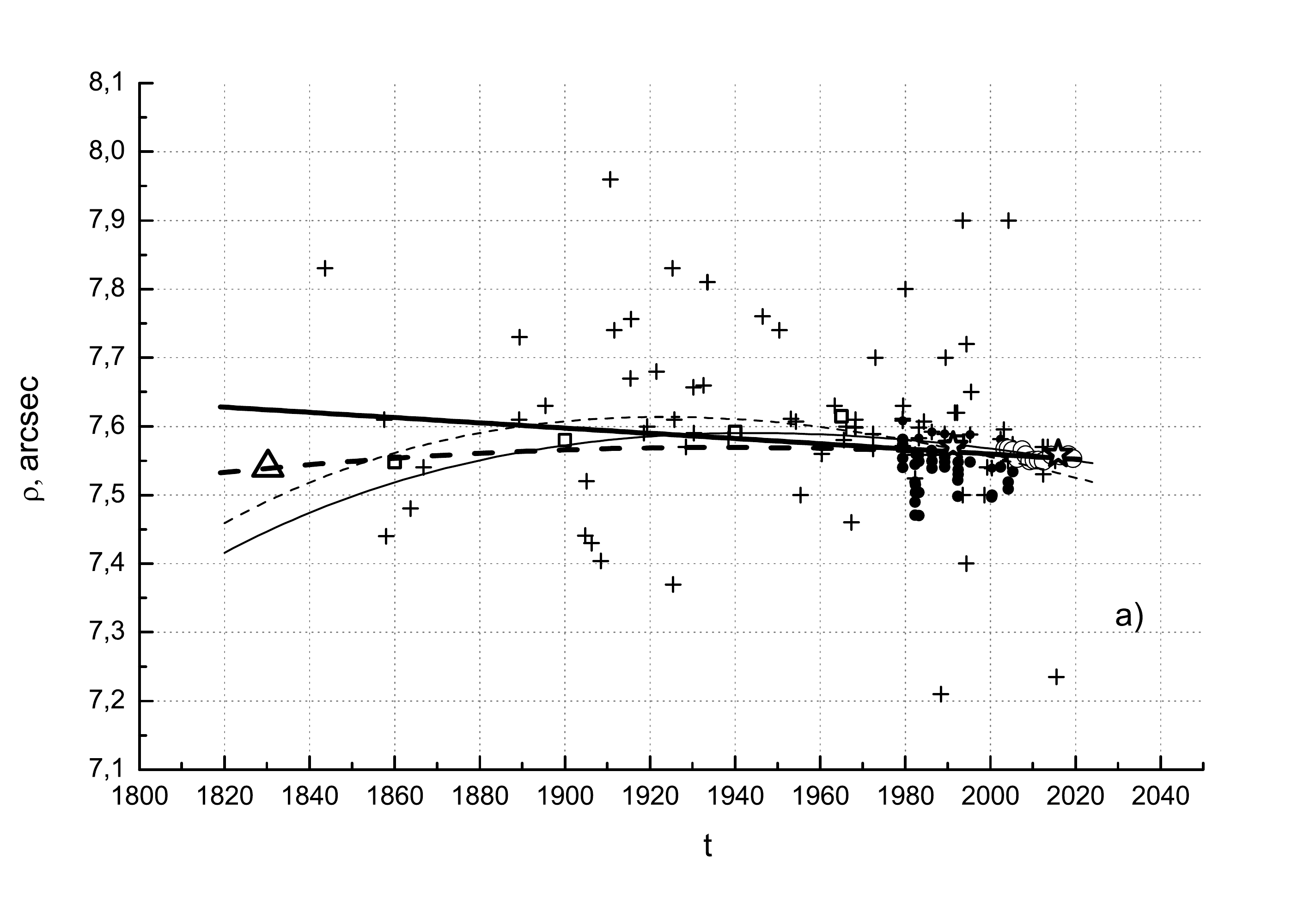}
   \includegraphics[width=0.7\textwidth]{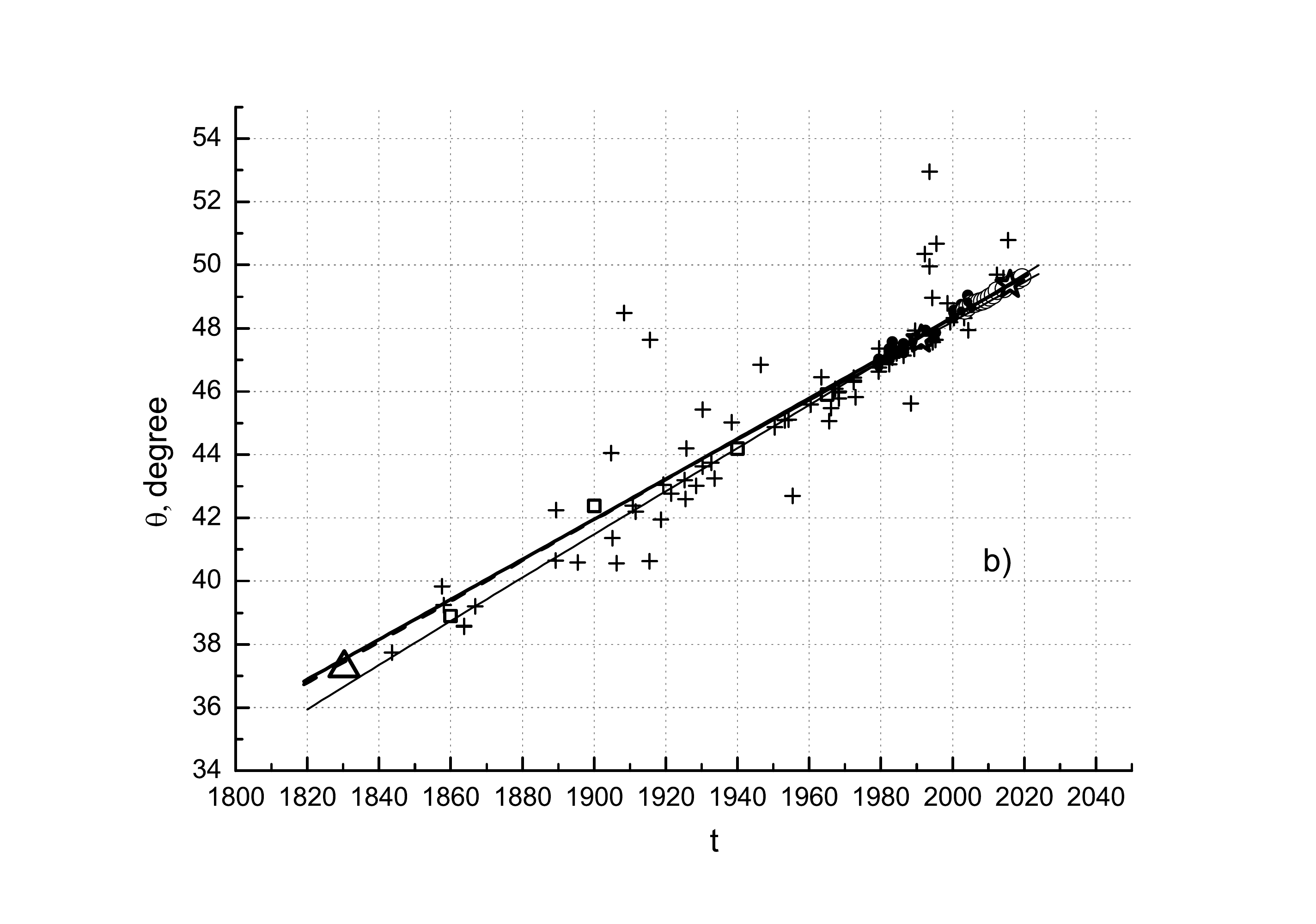}
   \includegraphics[width=0.7\textwidth]{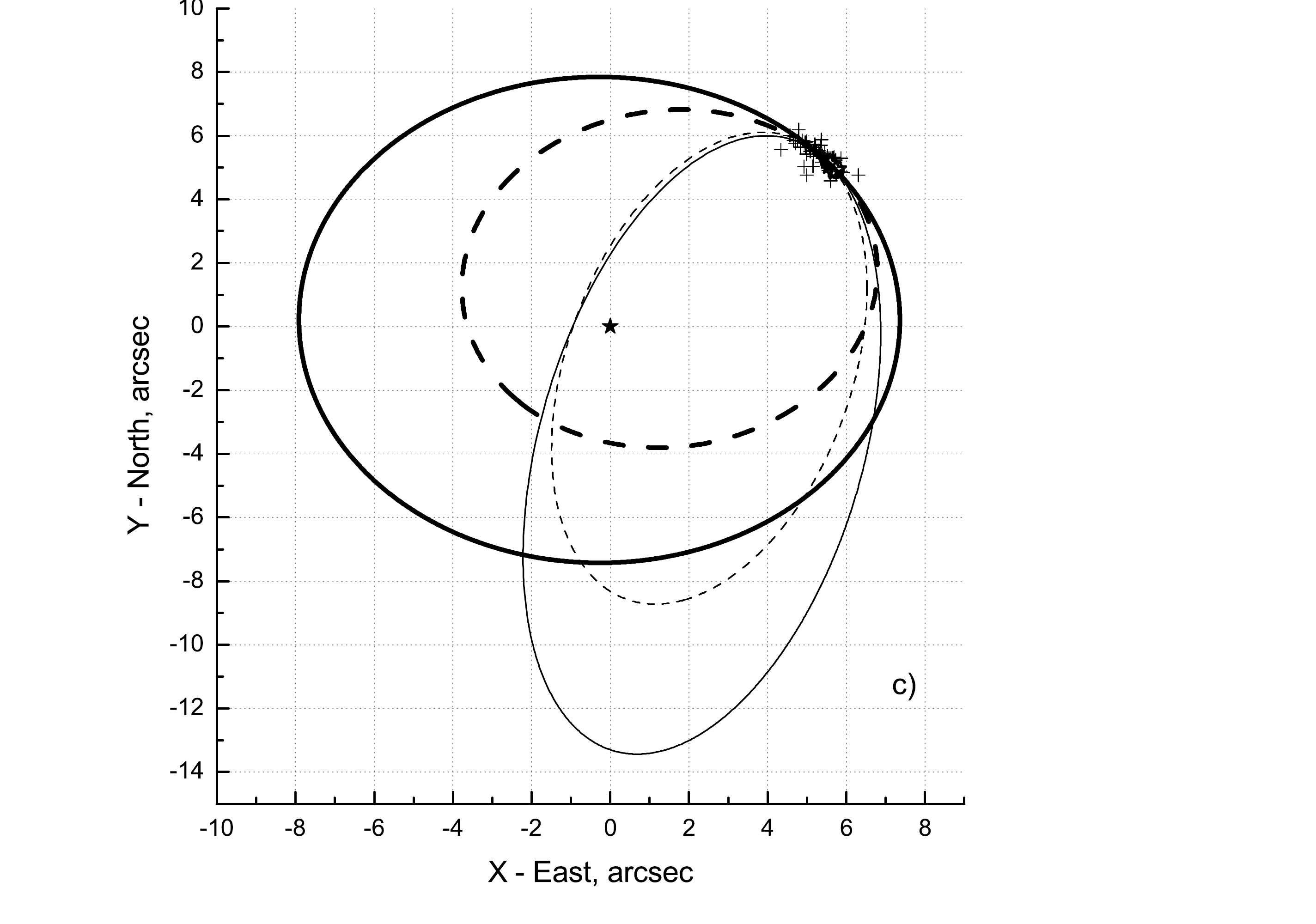}
   \caption{The comparison between the orbits of the external pair AB and observations. Crosses --- WDS observations, open squares --- the reference series, full and open circles --- the Pulkovo photographic and CCD observations, a triangle --- the first observation by Struve, stars --- the Hipparcos and Gaia~EDR3 observations, bold full line --- the orbit with 2.4$M_\odot$ (var.~2), bold dashes --- the orbit with 4.1$M_\odot$, thin full line --- the orbit from \cite{2019AstL...45...30I} with weights, thin dashes --- without weights.}
   \label{fig:out}
   \end{figure}

Since the length of a series of observations is longer than the orbital periods of a possible companion, the effect of its action is blurred and does not affect the calculation of the AMP. Taking our preliminary orbit into account does not improve the agreement with observations, but it does not distort the result either. The new orbits obtained in this work do not contradict the previous studies. Fig.~\ref{fig:out} presents observations and ephemerides of radically different orbits.

\section{Discussion}
\label{sect:discussion}

The question of the presence of the low-mass companion can be finally resolved only using the additional observations. Let us consider all the currently known arguments confirming the presence or absence of the low-mass companion.

\begin{figure} 
   \centering
   \includegraphics[width=\textwidth]{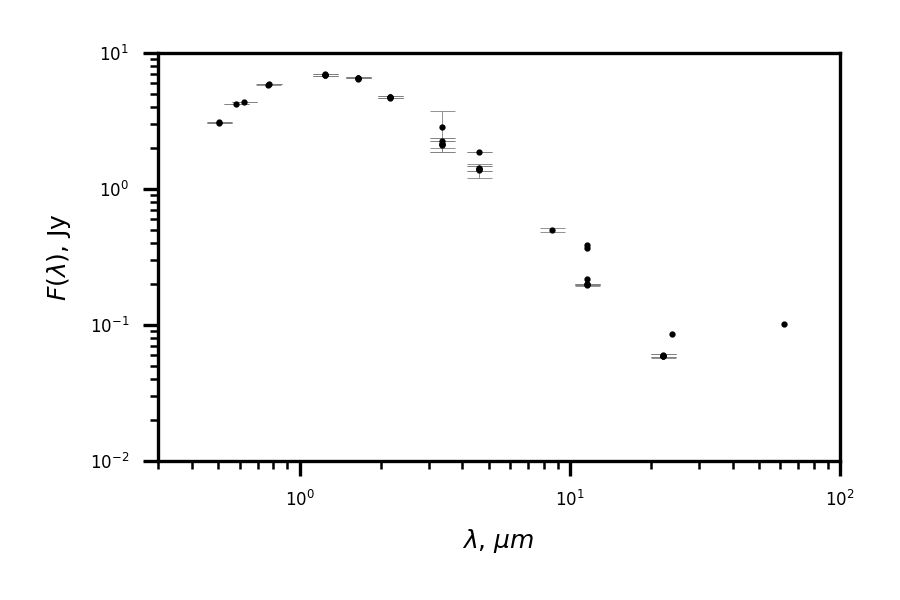}
   \includegraphics[width=\textwidth]{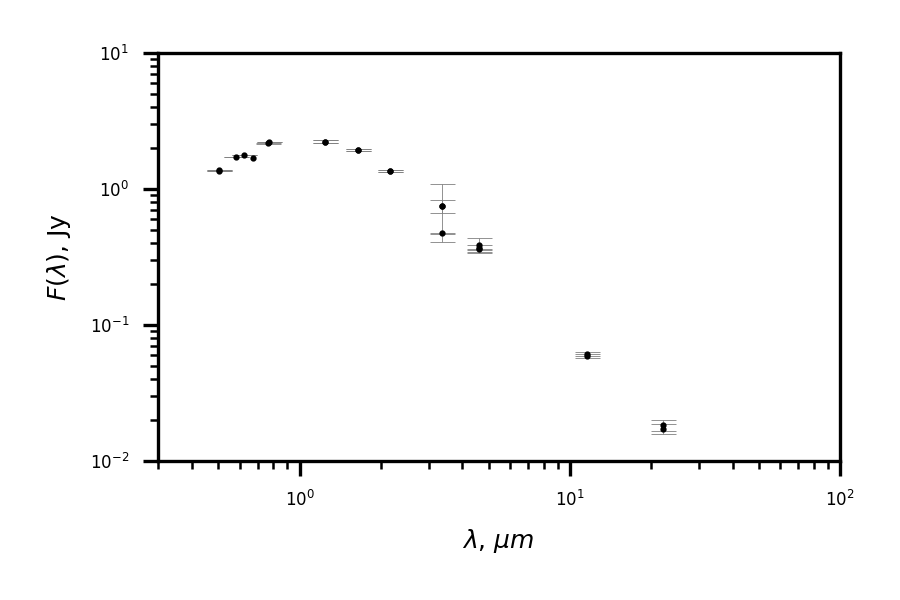}
   \caption{SED for the components of ADS~9346 (top --- for the A-star, bottom -- for the B-star). The data is from Gaia~DR2, Gaia~EDR3, 2MASS, WISE, IRAS.} 
   \label{fig:ir-excess}
   \end{figure}

\begin{enumerate}
    \item Signs of a companion for the A-star were obtained using the positional data but are in agreement with the independent radial velocity observations as well.
    
    \item In Gaia~EDR3 there is a parameter that characterizes the quality of astrometric observations, their agreement with models of the linear movement of the photocenter of a star, PSF and other effects. This parameter is astrometric\_excess\_noise\_sig or $D$ as in the original paper~(\citealt{2012A&A...538A..78L}).
    A significant exceeding of residual variance over the sum of squares of errors of individual effects ($D>2$) might be caused by different reasons or their combination. For instance, the images of bright stars are often overexposed. Another reason increasing the value $D$ -- the difference between the real movement of stars and a linear (5-parameter) model. In case of ADS~9346, $D=12.7$ for the A-star and $D=142.8$ for the B-star. An impact that a companion might have on total brightness with separation less than $0.2''$ is likely insignificant. Therefore, the reason why the parameter $D$ is large for ADS~9346 components might be associated with the overexposure, as the majority of stars surrounding ADS~9346 with magnitudes in the range from $7^m$ to $8.5^m$ are characterized by the parameter $D$ of the same order. Thus, we can conclude that the additional Gaia~EDR3 data limits the mass of a possible companion by 0.5~$M_\odot$. Otherwise, the parameter $D$ would be larger (the difference in magnitudes between the A-star and its companion would be $5^m$).
    \item Another argument confirming that the A-star has a companion follows from the analysis of the spectrum energy distribution (SED) based on the photometric observations conducted during the implementation of different observational programs in different spectrum ranges from Gaia to IRAS (Gaia~DR2, Gaia~EDR3, 2MASS~\citep{2006AJ....131.1163S}, WISE~\citep{2010AJ....140.1868W}, IRAS~\citep{2015A&C....10...99A}). SED for the components of ADS~9346 are presented in Fig.~\ref{fig:ir-excess}. As can be seen from the figure, there is an IR-excess for the A-star at 100~$\mu m$ (obtained by IRAS). According to the paper~(\citealt{2007ApJ...658.1289T}), this can be considered an indication of duplicity. This emission is generated by dust clouds near a relatively old star. The analysis performed by Trilling and coauthors~(\citealt{2007ApJ...658.1289T}) shows that stable large-scale dust structures near stars like the A-component of ADS~9346 can exist in regions of stable resonances in case of stellar multiplicity.
\end{enumerate}

\section{Conclusions}
\label{sect:conclusions}

The main result of our study is the confirmation of perturbation in the relative motion of A-B components of ADS~9346 based on observations conducted with the 26-inch Pulkovo refractor. The observed variations can be explained by the existence of a low-mass (0.129~$M_\odot$) companion that revolves around the A-star with the period of 15 years. The orbit of the companion is highly eccentric, the passage of the periastron in 2009 coincided with active observations with the 26-inch refractor, but there were no radial velocity observations that could confirm or disprove our conclusion. We calculated the preliminary orbit of the photocenter of the A-star plus the invisible companion consistent with the available radial velocity observations.

The presence of a companion is indirectly confirmed by the IR-excess in the emission of the A-star. For stars more than 5~Gy old dust radiation at a wavelength of 100~$\mu m$ can be caused by dust structures confined in regions of stable resonances.

Using the AMP method, we calculated new orbits of the external pair with a total mass given according to estimates based on photometric data and a mass obtained in best agreement with astrometric observations.

There are two possible reasons why these mass estimates differ:
\begin{itemize}
    \item The accuracy of astrometric observations collected over 200 years is low. The orbit that is in agreement with all observations can correspond to any mass (see Tab.~\ref{tab:orbout} and Fig.~\ref{fig:os_AB}). The Thiele-Innes elements can not be strictly determined on short arcs, therefore the formal best solution (4.1~$M_\odot$) is one of the possible solutions. However, it does contradict with other data. That is why we give two solutions: with the formally obtained mass and the expected mass.
    \item There is still a possibility that component B has another short-period companion which we can not detect according to our observations. Indirect evidence of such a scenario is that while there is a radial velocity measurement for component A in Gaia, there is none for component B; the parallax error of the component B in Gaia EDR3 is 3 times greater than that of the component A; as can be seen from Fig.~\ref{fig:cmd} the component B strongly moves away from the isochron to the blue side. All this gives reason to believe that the properties of this star do not fit into the formal model.
\end{itemize}

The AMP method allows the consistent use of the object observation results obtained by different methods, and therefore leads to more reliable results, which is confirmed by a comparison with the orbits published in ~\cite{2019AstL...45...30I}. For high-volume determination of orbits for the purpose of statistical studies, formal methods are acceptable, but on short arcs they cannot lead to a reliable result. For the study of individual stars, the AMP method is preferable, supplemented by agreement with all available photometric data and radial velocity measurements.

\normalem
\begin{acknowledgements}
The reported study was funded by RFBR according to the research projects 19-02-00843~A and 20-02-00563~A and with partial support of the grant 075-15-2020-780 of the Government of the Russian Federation and the Ministry of Higher Education and Science. 

This work has made use of data from the European Space Agency (ESA) mission {\it Gaia} (\url{https://www.cosmos.esa.int/gaia}), processed by the {\it Gaia} Data Processing and Analysis Consortium (DPAC, \url{https://www.cosmos.esa.int/web/gaia/dpac/consortium}). Funding for the DPAC has been provided by national institutions, in particular the institutions participating in the {\it Gaia} Multilateral Agreement.

\end{acknowledgements}
  
\bibliographystyle{raa}
\bibliography{kiyaeva}

\end{document}